\documentclass[prb,reprint,citeautoscript,showpacs]{revtex4-1}
\usepackage{graphicx}
\usepackage{amssymb}
\usepackage{amsmath}
\usepackage{url}
\usepackage{prettyref}
	
	\newrefformat{app}{appendix~\ref{#1}}
	\newrefformat{eqn}{Eqn.~(\ref{#1})}
	\newrefformat{eqns}{Eqns.~(\ref{#1})}
	\newrefformat{sec}{section~\ref{#1}}
	\newrefformat{subsec}{section~\ref{#1}}
	\newrefformat{subsubsec}{section~\ref{#1}}
	\newrefformat{fig}{Fig.~\ref{#1}}
\usepackage[usenames,dvipsnames]{color}
\usepackage[breaklinks,bookmarksnumbered,colorlinks=true,citecolor=blue,urlcolor=blue,linkcolor=red]{hyperref}
\begin{document}
\title{Flexural mode of graphene on a substrate}
\author{Bruno Amorim}
\email[E-mail me at: ]{amorim.bac@icmm.csic.es}
\affiliation{%
Instituto de Ciencia de Materiales de Madrid,
CSIC,
Cantoblanco,
E-28049, Madrid, Spain
}
\author{Francisco Guinea}
\affiliation{%
Instituto de Ciencia de Materiales de Madrid,
CSIC,
Cantoblanco,
E-28049, Madrid, Spain
}
%
%
%
%
%
%
%
%
%
\begin{abstract}
Out of plane vibrations are suppressed in graphene layers placed on a substrate. These vibrations, in suspended samples, are relevant for the understanding of properties such as the electrical resistivity, the thermal expansion coefficient, and other. We use a general framework to study the properties of the out of plane mode in graphene on different substrates, taking into account the dynamics of the substrate. We discuss broadening of this mode and how it hybridizes with the substrate Rayleigh mode, comparing our model with experimental observations. We use the model to estimate the substrate induced changes in the thermal expansion coefficient and in the temperature dependence of the electrical resistivity.
\end{abstract}
\maketitle
%
%
%
\section{Introduction}
Since its isolation in 2004 \cite{Novoselov_2004}, graphene, a monolayer
of carbon atoms arranged in a honeycomb lattice, has received great
attention due to both its unique electrical and mechanical properties \cite{RevModPhys_2009}.
In graphene, the carbon atoms display an sp$^{2}$ hybridization,
with the out of plane p$_{z}$ orbitals forming a $\pi$ band which
is responsible for its electrical properties, while the sp$^{2}$
orbitals form strong $\sigma$ in plane bonds that govern its mechanical
properties. It has been verified both experimentally and theoretically \cite{Mounet_2005,Lee_2008,Zakharchenko_2009} that graphene is the known material with the largest in plane elastic
constants.

In freely suspended graphene samples the vibrations of the lattice can be classified into in plane and out of plane (flexural) modes, with the flexural mode lying at lower energies and showing a quadratic dispersion. Anharmonic effects at long wavelengths strongly couple in plane and flexural modes \cite{Nelson_Membranes}. The flexural mode is responsible for the significant temperature dependence of the electronic resistivity at low temperatures~\cite{Mariani_2008,Mariani_2010,EdCastro_2010,HOchoa_2011,HOchoa_2012}. Anharmonic effects in suspended graphene explain its negative thermal expansion coefficient \cite{Paco_2012} and they play a significant role in the thermal conductivity~\cite{Ong_2011}.

The interaction between a graphene layer and a substrate underneath changes significantly the properties of the out of plane vibrations of the entire system~\cite{Aizawa_1990PRL,Aizawa_1990PRB,Ong_2011,Qiu_2012}. Coupling to a substrate also leads to heat transfer between the two systems~\cite{Persson_2010,Persson_2010b,Persson_2011}.

The deformations of the hybrid system made up of the graphene layer and the substrate at small amplitudes and long wavelengths are rigorously described by the theory of elasticity.
This theory fixes the number of independent couplings required, which is determined by the dimensionality and symmetries of the two systems, graphene and substrate, to be studied. A model fully consistent with the theory of elasticity is described in the next section.
Being general, this model should also describe other two dimensional materials supported by a substrate. We studied how this coupling gives origin to a finite lifetime for the flexural mode and to a hybridization of this with the substrate surface Rayleigh mode, comparing our model with experimental data from Ref.~\onlinecite{Aizawa_1990PRB}. We also studied the thermal expansion of graphene on a substrate and also the effect of the flexural mode,
modified by the coupling to the substrate, on the electrical resistivity of doped graphene, focusing on two of the most common substrates: silicon dioxide, SiO$_{2}$, and
hexagonal boron nitride, hBN.

\begin{figure}
\begin{centering}
\includegraphics[width=3.5cm]{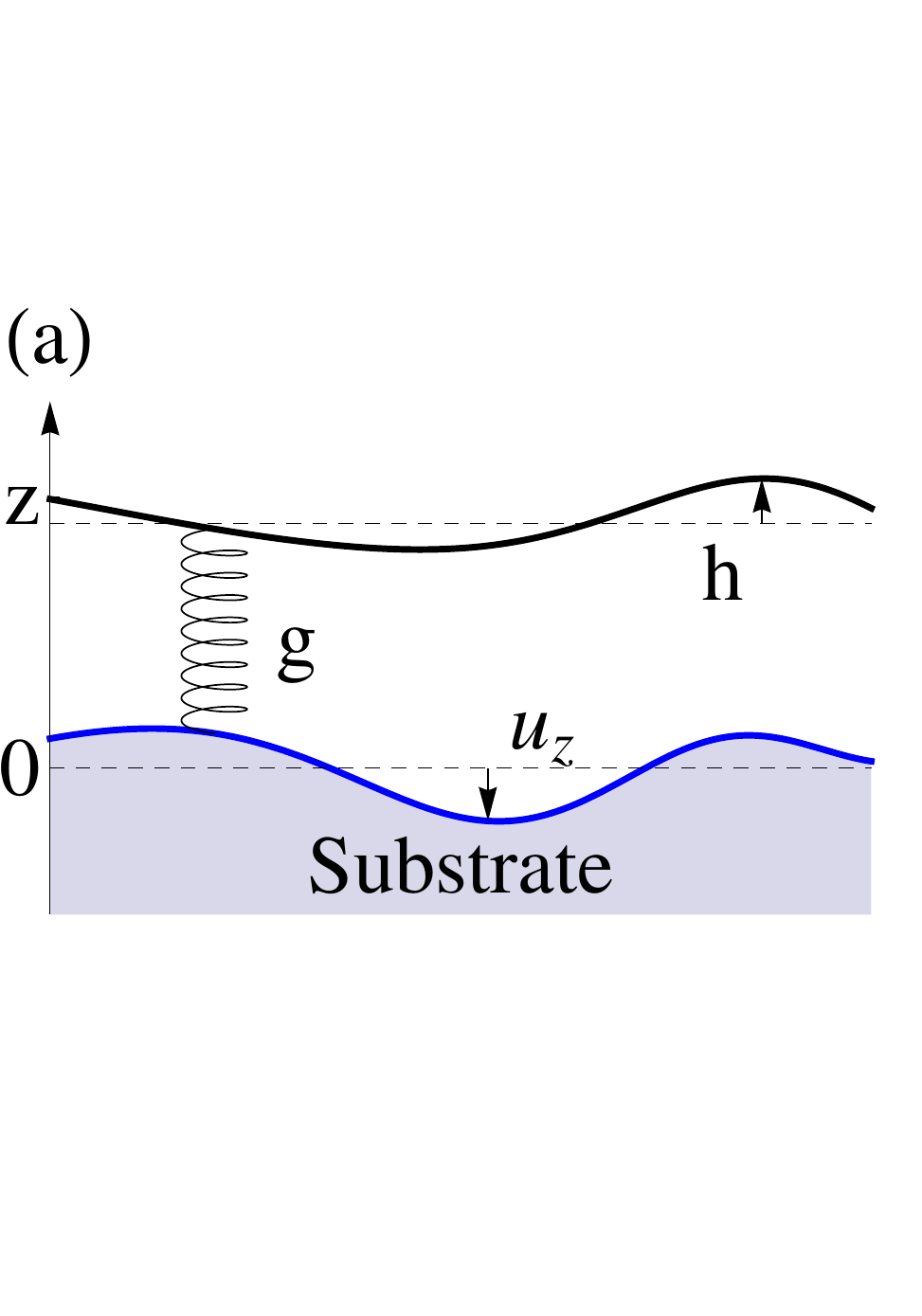}\includegraphics[width=5cm]{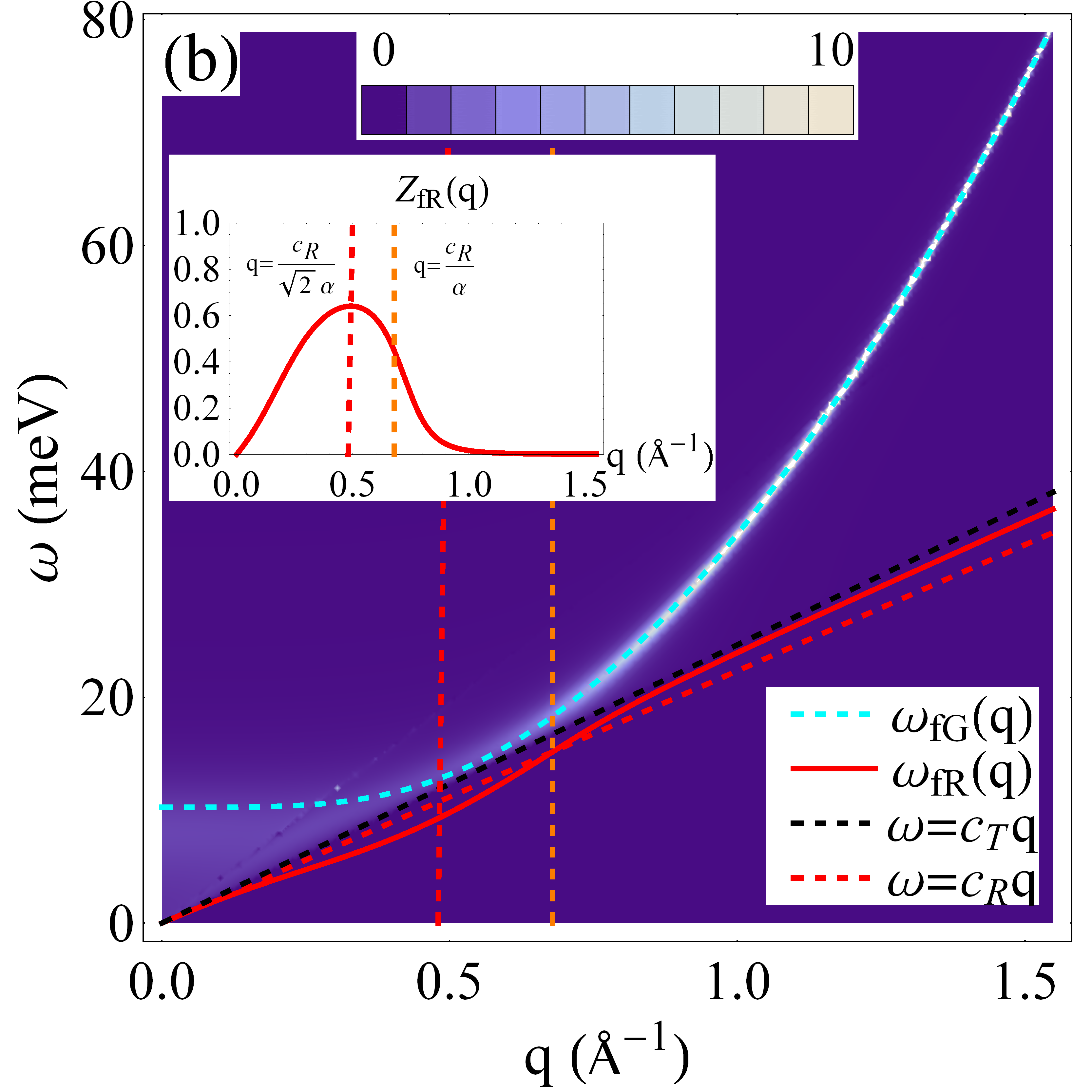}
\par\end{centering}

\caption{\label{fig:Model}(a) Representation of the model used for
the graphene membrane-substrate coupling. (b) Density plot of the spectral
function $A(q,\omega)$ for graphene on SiO$_{2}$ [in units of $\pi\omega_{0}^{2}/(2\gamma_{0})$]. For $\omega<c_{\text{T}}q$, it is zero everywhere, except at the dispersion relation of the fR mode, $\omega_{\text{fR}}(q)$,  where it is a Dirac delta function with weight $Z_{\text{fR}}(q)$ shown in the inset. For $\omega>c_{\text{T}}q$, $A(q,\omega)$ is finite, with a peak close to $\omega_{\text{fG}}(q)=\sqrt{\alpha^{2}q^{4}+\omega_{0}^{2}}$, which becomes very broad for small $q$, indicating that the flexural phonon becomes poorly defined. Vertical lines are $q=c_{\text{R}}/(\sqrt{2}\alpha)$ and $q=c_{\text{R}}/\alpha$. We used $g=1.82\times10^{20}\text{J}/\text{m}^{4}$.
}
\end{figure}

\section{The model}
\label{sec:Model}
In our model, a flat graphene membrane is supported by a semi-infinite flat substrate\footnote{
If the substrate is not flat, such as SiO$_2$,
graphene will have regions where it conforms to the substrate, while other regions will become detached from it~\protect \cite{Paco_2011}. The detailed analysis of this situation is beyond the scope of the present paper, although the analysis presented provides a reasonable approximation to the regions which are well attached to the substrate.
}
 that occupies the half-space $z<0$. In a long wavelength description, we will use the elastic theory of a crystalline membrane to model graphene and linear elasticity theory to describe the substrate. Therefore, the action describing the membrane-substrate coupled system will be given by $S=S_{\text{out}}+S_{\text{in}}+S_{\text{subs}}+S_{\text{coup}}+S'_{\text{coup}}$. $S_{\text{out}}$ is the quadratic action for the flexural mode,  $S_{\text{in}}$ is the action for the in plane modes of the membrane including anharmonic coupling between in plane and flexural modes,  $S_{\text{subs}}$ is the linear elasticity action for the substrate and  $S_{\text{coup}}$ and  $S'_{\text{coup}}$ describe the membrane-substrate coupling. Assuming that the fluctuations around the equilibrium positions are small, we keep $S_{\text{coup}}$ and $S'_{\text{coup}}$ only to quadratic order in the displacement fields\cite{Andelman_1999}.

Finally, assuming in plane isotropy, the most general description of the membrane-substrate model consistent with the theory of elasticity is
\begin{align}
S_{\text{out}} & = \frac{1}{2}\int dtd^{2}x\left(\rho_{\text{2D}}\dot{h}^{2}-\kappa\left(\partial^{2}h\right)^{2}\right),\nonumber \\
S_{\text{in}} & = \frac{1}{2}\int dtd^{2}x\left(\rho_{\text{2D}}\dot{\vec{u}}^{2}-\lambda\varepsilon_{\alpha\alpha}^{2}-2\mu\varepsilon_{\alpha\beta}\varepsilon_{\alpha\beta}\right),\nonumber\\
S_{\text{subs}} & = \frac{1}{2}\int_{z<0}dtd^{3}x\left(\rho_{\text{3D}}(\dot{\vec{u}}^{(s)})^{2}-c_{ijkl}\partial_{i}u_{j}^{(s)}\partial_{k}u_{l}^{(s)}\right),\nonumber \\
S_{\text{coup}} & =-\frac{g}{2}\int_{z=0}dtd^{2}x\left(h-u_{z}^{(s)}\right)^{2}, \nonumber \\
S'_{\text{coup}} & = -\frac{g'}{2}\int_{z=0}dtd^{2}x\left(\frac{h-u_{z}^{(s)}}{d}\right)\left(\partial_{\alpha}u_{\alpha}+\partial_{\alpha}u_{\alpha}^{(s)}\right)\nonumber\\
    & -\frac{g''}{2}\int_{z=0}dtd^{2}x\sum_{\alpha}\left(\frac{u_{\alpha}-u_{\alpha}^{(s)}}{d}+\frac{\partial_{\alpha}h+\partial_{\alpha}u_{z}^{(s)}}{2}\right)^{2}.
\end{align}
Here $h$ and $u_{\alpha}$ are, respectively, the out of plane and in plane displacement fields of the membrane with mass density per unit area $\rho_{\text{2D}}$, bending rigidity $\kappa$ and Lam\'{e} coefficients $\lambda$ and $\mu$. $\varepsilon_{\alpha\beta}=\left(\partial_{\alpha}u_{\beta}+\partial_{\beta}u_{\alpha}+\partial_{\alpha}h\partial_{\beta}h\right)/2$
is the strain tensor to lowest order in $h$ and $u_{\alpha}$. $u_{i}^{(s)}$ is the displacement field of the substrate with mass density per volume $\rho_{\text{3D}}$ and elastic constants $c_{ijkl}$ (see Table~\ref{tab:material}) (greek indices run from 1 to 2 and latin indices from 1 to 3). For graphene, $\rho_{\text{2D}}=7.6\times10^{-8}\text{g}/\text{cm}^2$ and $\kappa\approx1.1\text{eV}$ \cite{Zakharchenko_2009}.

The graphene-substrate coupling $g$ in $S_{\text{coup}}$ is determined by the van der Waals interaction between the two systems. The couplings in $S'_{\text{coup}}$ depend on short range interatomic interactions, and are significantly weaker than the van der Waals interaction for graphite (graphene on graphene)\cite{Tetal12}. Out of plane phonons of graphene are also the most affected by the presence of a substrate \cite{Aizawa_1990PRL,Aizawa_1990PRB,Aizawa_1990Ni,Aizawa_1992,Yanagisawa_2005,Wirtz_2010,Politano_2012}. Having this in mind, in the following we will ignore $S'_{\text{coup}}$.

We expect that intrinsic anharmonic effects of the membrane, important in a free standing membrane \cite{Nelson_Membranes}, will be unimportant in the presence of a substrate and we will therefore ignore them unless when discussing thermal expansion. In this approximation, in plane and out of plane modes decouple and we can ignore the term $S_{\text{in}}$, leaving us with a harmonic theory.\footnote{
This same model was previously used to describe the sliding friction of rubber \protect \cite{Persson_2001} and heat transfer between a membrane and a substrate \protect \cite{Persson_2010,Persson_2010b,Persson_2011}. It is also a continuous version of the force constant model used in Ref.~\protect \onlinecite{Aizawa_1990PRB}, but taking into account the dynamics of the substrate.
}
For a free standing membrane, $S_{\text{out}}$ gives a quadratic dispersion relation $\omega_{\text{f}}(q)=\alpha q^{2}$, $\alpha=\sqrt{\kappa/\rho_{\text{2D}}}$. Coupling to a static substrate will gap this dispersion relation and one would obtain $\omega_{\text{fG}}(q)=\sqrt{\alpha^{2}q^4 + \omega_{0}^{2}}$, with $\omega_{0}=\sqrt{g/\rho_{\text{2D}}}$.

The value of the constant $g$ greatly varies from substrate to substrate (see Table~\ref{tab:material}). It was estimated in Ref.~\onlinecite{Persson_2010} to have a value of $1.82\times10^{20}\text{J}/\text{m}^{4}$ for graphene on SiO$_{2}$. For graphene on hBN its value can be estimated from density functional theory (DFT) calculations \cite{Giovannetti_2007} to be around $1.2-2.7\times 10^{20}\text{J}/\text{m}^4$, depending on the orientation of graphene on hBN. For the (111) surface of transition metal carbides it is of the order of $2\times10^{21}\text{J}/\text{m}^4$, while for the (001) face it is approximately zero \cite{Aizawa_1990PRB}, as it is for graphene on platinum (111) \cite{Aizawa_1992,Politano_2012}.

The main object of interest from which all physically relevant quantities can be obtained is the
height-height retarded Green's function
\begin{equation}
\label{Greens}
D(q,\omega)=-\frac{i}{\hbar}\int dtd^{2}\Theta(t)xe^{i(\omega t-\vec{q}\cdot\vec{x})}\left\langle \left[h(x,t),h(0,0)\right]\right\rangle,
\end{equation}
where $\left\langle \right\rangle $ means thermal and quantum average.
Ignoring $S_{\text{in}}$ it is possible to solve
the theory exactly. As a matter of fact, the problem reduces to that of two
coupled harmonic oscillators. Therefore one obtains
\begin{equation}
D(q,\omega)=\left[\rho_{\text{2D}}\omega^{2}-\kappa q^{4}-\Pi(q,\omega) \right]^{-1},
\end{equation}
where $\Pi(q,\omega)=g\left[  1-g\Delta_{0}(q,\omega) \right] ^{-1}$ is the correction to the free propagator due
to coupling to the substrate, with $\Delta_{0}(q,\omega)$ the surface-to-surface substrate
propagator  similarly defined as in Eq.~\ref{Greens}, with both fields $u_{z}^{(s)}$ evaluated at the surface of the substrate, $z=0$, and the index $_{0}$ meaning $g=0$. Since $S_{\text{subs}}$ is quadratic, to obtain
$\Delta_{0}(q,\omega)$ it suffices to study the classical
response of the substrate to an external pressure at the boundary
$z=0$. This was done for an isotropic substrate in Ref.~\onlinecite{Persson_2001} and we generalized the result for the case of an uniaxial substrate (see Appendix~\ref{appendix} for
details). There are two specially relevant cases for the behaviour of $\Delta_{0}(q,\omega)$: $q=0$, for which $\Delta_{0}(0,\omega)=-i/(\omega c_{\text{L}} \rho_{\text{3D}})$ (replace $c_{\text{L}}\rightarrow\sqrt{c_{33}/\rho_{\text{3D}}}$ for the uniaxial case); and $\omega=0$, for which $\Delta_{0}(q,0)=-1/(K_{1}q)$, with $K_{1}=2\rho_{\text{3D}}c_{\text{T}}^{2}\left(c_{\text{L}}^{2}-c_{\text{T}}^{2}\right)/c_{\text{L}}^{2}$
for an isotropic medium (where $c_{\text{T/L}}$ is the transverse/longitudinal sound
velocity of the substrate). Therefore, at small $q$, coupling to the substrate will lead to a contribution to $D(q,0)^{-1}$ proportional to $q$, while the first order contribution to $D(q,0)^{-1}$ arising from intrinsic anharmonic effects in a free standing membrane is proportional to $q^{2}$\cite{Nelson_Membranes}. Therefore, anharmonic effects will be irrelevant when comparing to the effect of the substrate. This justifies our approximation of neglecting $S_{\text{in}}$. 

A semi-infinite
elastic medium supports a continuum of 3D bulk modes for
$\omega>c_{\text{T}}q$ (replace $c_{\text{T}}\rightarrow\sqrt{c_{44}/\rho_{\text{3D}}}$
for the uniaxial case). For $\omega<c_{\text{T}}q$, the substrate
supports a 2D surface Rayleigh
mode with dispersion given by $\omega=c_{\text{R}}q$, with $c_{\text{R}}$ the Rayleigh velocity. Therefore the effect
of coupling to the substrate on the flexural mode will be twofold: (i) Coupling to the substrate will gap the dispersion relation of flexural mode, $\omega_{\text{fG}}(q)$, which we will refer to as the flexural-gapped (fG) mode, so that it will in general lie within the continuum of substrate bulk modes, which act as a dissipative bath, leading to a broadening of this branch. (ii) The flexural mode will also hybridize with the Rayleigh mode
(already pointed out in Ref.~\onlinecite{Ong_2011}) giving origin to another, unbroadened,
branch, $\omega_{\text{fR}}(q)$, which we will refer to as the flexural-Rayleigh (fR) mode. This information is encoded in the spectral/dissipation function, defined as\footnote{
This quantity obeys the sum rule $\protect\int_{0}^{\infty} d \omega A(q,\omega )=1 $, as a consequence of the canonical commutation relation $\left [h(x,t),\pi (y,t)\right ]=i\hbar \delta(x-y)$, where $\pi (x,t)=\rho _{\text {2D}}\dot{h}(x,t)$
} 
\begin{equation}
A(q,\omega)=-\frac{{2\rho_{\text{2D}}}\omega}{\pi}\text{Im}D(q,\omega).
\end{equation}
The fR mode appears in $A(q,\omega)$ as a Dirac delta function divergence at $\omega=\omega_{fR}(q)$, with $\omega_{\text{fR}}(q)$ the solution of $D(q,\omega_{\text{fR}}(q))^{-1}=0$ for $\omega<c_{\text{T}}q$, with a weight $Z_{\text{fR}}(q)$ given by 
\begin{equation}
Z_{\text{fR}}(q)^{-1}=1-\frac{1}{2\rho_{\text{2D}}\omega}\frac{\partial}{\partial\omega}\text{Re}\Pi(q,\omega)\biggr|_{\omega=\omega_{\text{fR}}(q)}.
\end{equation} 
This situation is illustrated in Fig.~\ref{fig:Model}b, where
we show a density plot of the spectral function along with the dispersion
of the fR mode. The gap of the fG mode is controlled by $\omega_{0}=\sqrt{g/\rho_{\text{2D}}}$, while the broadening is controlled by $\gamma_{0}=g/(c_{\text{L}}\rho_{\text{3D}})$ (replace $c_{\text{L}}\rightarrow\sqrt{c_{33}/\rho_{\text{3D}}}$ for the uniaxial case).
For $g=1.82\times10^{20}\text{J}/\text{m}^{4}$ one obtains $\omega_{0}\approx10\text{ meV}$
and $\gamma_{0}\approx13\text{ THz}$ for graphene on SiO$_{2}$. This value
is an overestimation comparing with the inverse relaxation times
obtained from molecular dynamics simulations for acoustic flexural
phonons on a SiO$_{2}$ substrate, $1/\tau\sim0.1-1\text{ THz}$~\cite{Qiu_2012}.
The hybridization between flexural and Rayleigh modes is more relevant for values of the spring
constant $g$ such that $\omega_{0}\sim c_{\text{R}}^{2}/(2\alpha)$, being maximum in this situation for $q\sim c_{\text{R}}/(\sqrt{2}\alpha)$ and being suppressed  for $q\gtrsim c_{\text{R}}/\alpha$.

\begin{figure*}
\begin{centering}
\includegraphics[height=5.5cm]{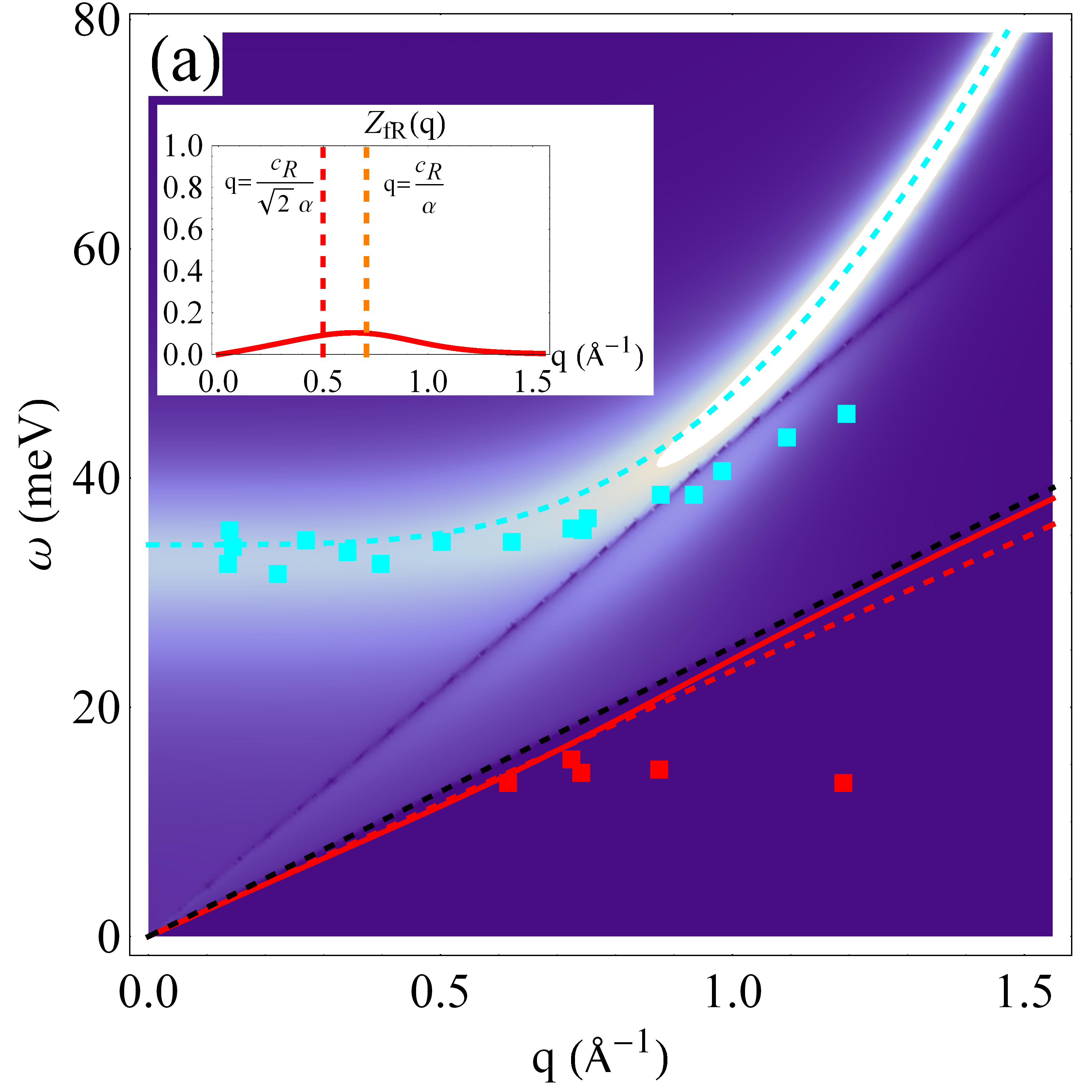}\includegraphics[height=5.5cm]{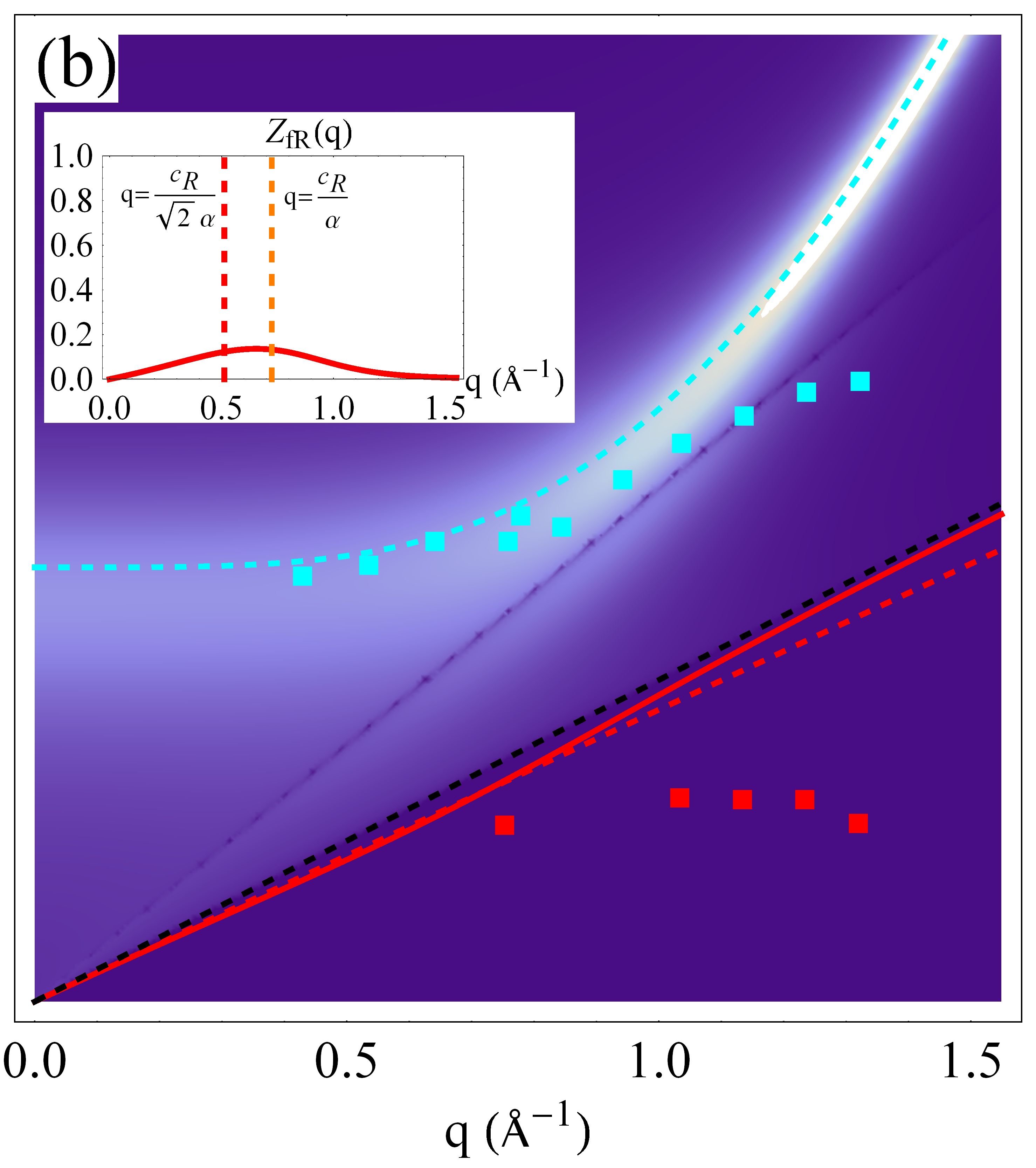}\includegraphics[height=5.5cm]{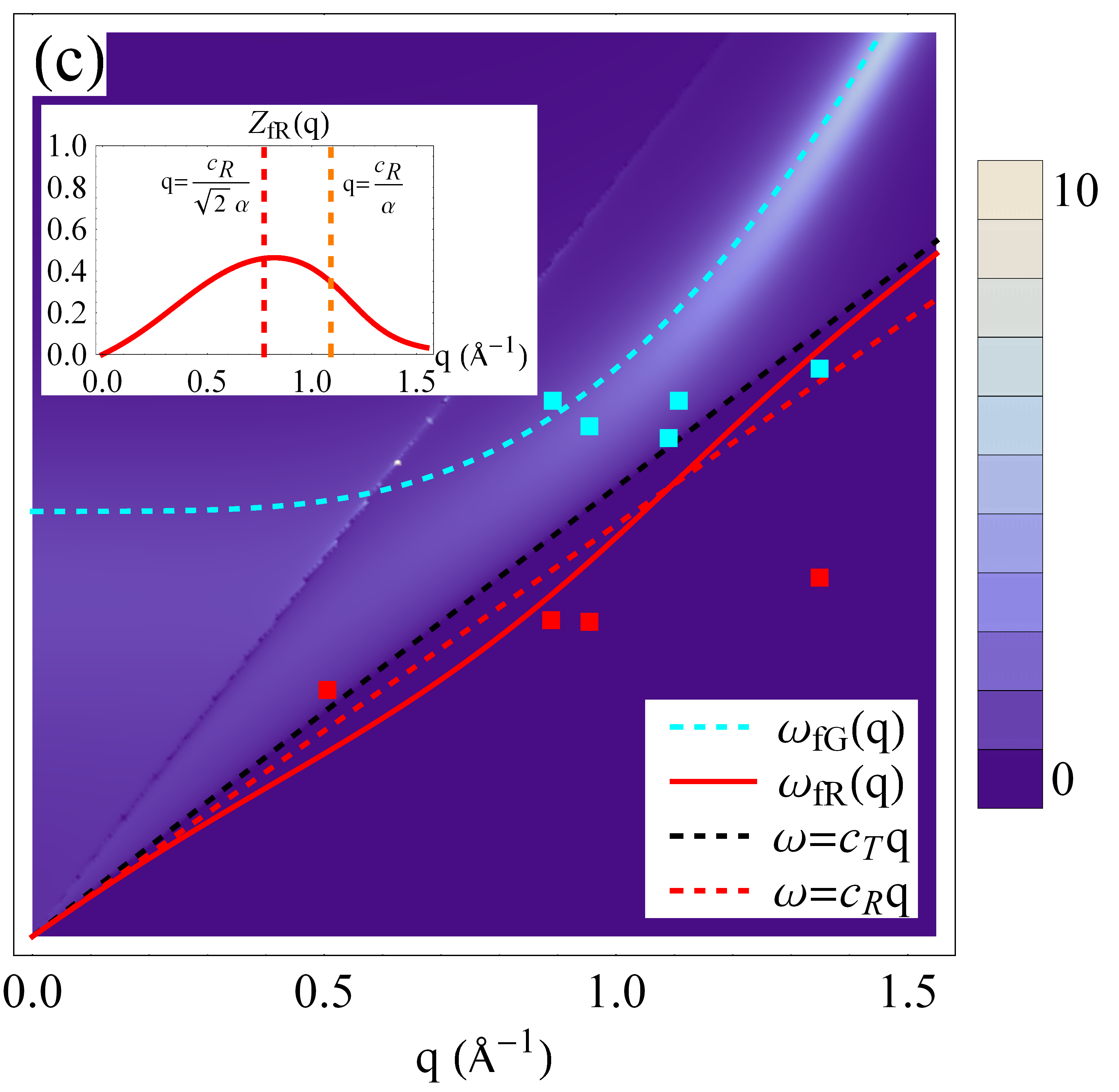}
\par\end{centering}

\caption{\label{fig:exp}Comparison of computed $A(q,\omega)$ [density plot in units of $\pi\omega_{0}^{2}/(2\gamma_{0})$] with experimental phonon dispersion relations (solid squares) for graphene on three substrates, (a) TaC, (b) HfC and (c) TiC, obtained via high resolution electron energy loss spectroscopy (HREELS) in Ref.~\onlinecite{Aizawa_1990PRB} (cyan squares: flexural mode; red squares: possible Rayleigh mode). Insets: weight of fR mode on the graphene membrane.
}

\end{figure*}

\begin{table*}
\begin{centering}
\begin{tabular}{ c c c c c c c c c c c }
\hline\hline
 & $\rho_{\text{3D}}$(g cm$^{-3}$) & $c_{11}$ (GPa) & $c_{12}$ & $c_{13}$ & $c_{33}$ & $c_{44}$ & $g$ ($10^{20}$N m$^{-3}$) & $\omega_{0}$ (meV) & $\gamma_{0}$ & $c_{R}$ (m s$^{-1}$)\tabularnewline
\hline
SiO$_{2}$ & 2.20 & 78 & - & - & - & 31 & 1.82\cite{Persson_2010} & 10 & 9 & 3392\tabularnewline

hBN  & 2.28 & 811\cite{Bosak_2006} & 169\cite{Bosak_2006} & 0\cite{Bosak_2006} & 27\cite{Bosak_2006} & 7.7\cite{Bosak_2006} & 1.2-2.7\cite{Giovannetti_2007} & 10 & 15 & 1835\\

TaC  & 14.65 & 634\cite{Brown_1966} & - & - & - & 216\cite{Brown_1966} & 20.23\cite{Aizawa_1990PRB} & 34 & 14 & 3525\\

HfC  & 12.27 & 500\cite{Brown_1966} & - & - & - & 195\cite{Brown_1966} & 21.72\cite{Aizawa_1990PRB} & 35 & 18 & 3681\\

TiC  & 4.94 & 500\cite{Gilman_1961} & 113\cite{Gilman_1961} & - & - & 175\cite{Gilman_1961} & 23.82\cite{Aizawa_1990PRB} & 37 & 32 & 5453\\

\hline\hline
\end{tabular}
\par\end{centering}

\caption{\label{tab:material}Material parameters for different substrates and computed values for
$\omega_{0}$, $\gamma_{0}$ and $c_{\text{R}}$. The transition metal carbides were approximated by isotropic materials, with the data for TaC and HfC taken from polycrystalline samples (Ref.~\onlinecite{Brown_1966}), while for TiC only the constants $c_{11}$ and $c_{44}$ were used.}
\end{table*}

\begin{figure}
\begin{centering}
\includegraphics[width=6.5cm]{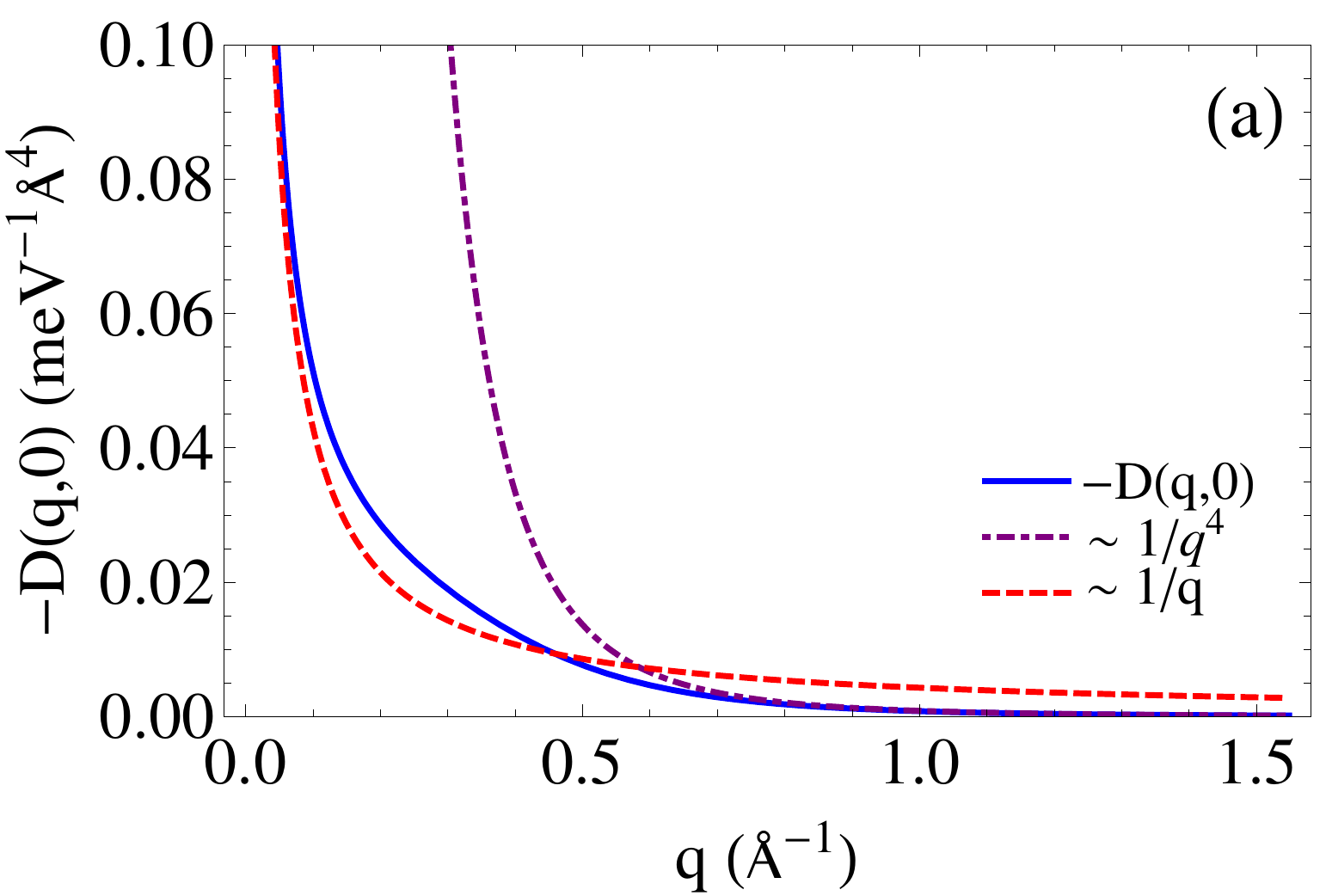}
\includegraphics[width=6.5cm]{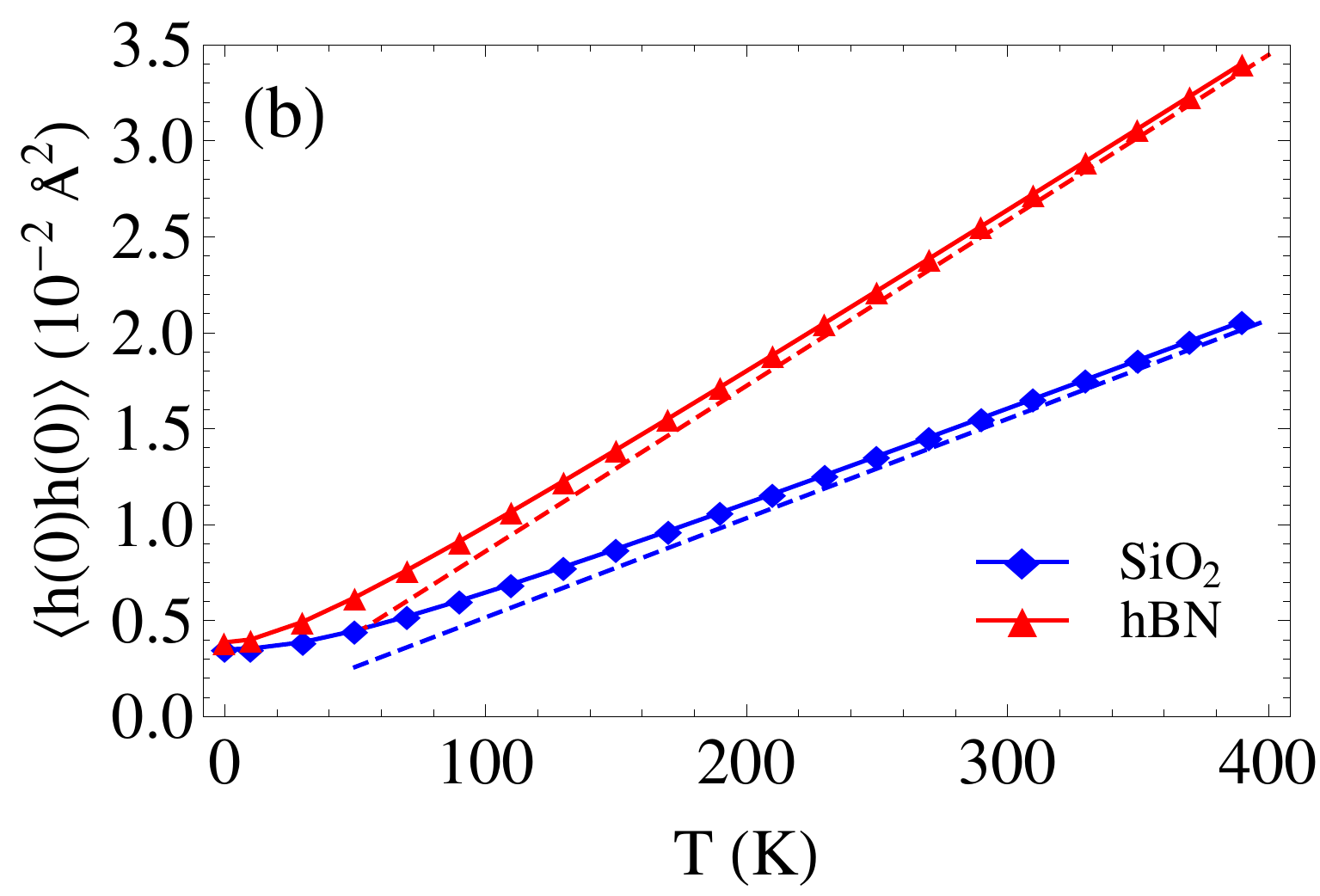}
\par\end{centering}

\caption{\label{fig:height-height}(a) Plot of $D(q,0)$ for graphene on a SiO$_{2}$ substrate.
Its behaviour changes from the one of a free membrane at large $q$, $1/q^{4}$, to the one of the substrate at low $q$, $1/q$. (b) Plot of $\left\langle h(0)h(0)\right\rangle $
as a function of temperature for SiO$_{2}$ and hBN substrates (solid lines) and high temperature limits (dashed lines). $g=1.82\times10^{20}\text{J}/\text{m}^{4}$ was used for both substrates.
}
\end{figure}

\section{Results}

\subsection{Comparison with experimental results}
It is interesting to compare our model with experimental data from Ref.~\onlinecite{Aizawa_1990PRB} of phonon dispersion relations of graphene on different substrates. This is shown in Fig.~\ref{fig:exp}. Although our continuous model fails at large momenta, it semi-quantitatively explains the lack of experimental data for the flexural mode at low momenta for graphene on light substrates, since the phonons become ill defined as quasi-particles. Also notice that experimentally there are indications of a Rayleigh mode. Our model predicts that just by probing the carbon layer it is possible to detect the hybrid fR mode. If this is the case or if what is experimentally seen comes from the fact that the first few layers of the substrate are also being probed is not clear.

\subsection{Height-height correlation function}
The equal time height-height correlation
function is given by
\begin{equation}
\left\langle h(x)h(0)\right\rangle =-\hbar\int\frac{d^{2}qd\omega}{\left(2\pi\right)^{3}}e^{i\vec{q}\cdot\vec{x}}\coth\left(\frac{\hbar\omega}{2k_{B}T}\right)\text{Im}D(q,\omega),
\end{equation}
which at high temperature reduces to $\left\langle h(x)h(0)\right\rangle \simeq-k_{B}T\int\frac{d^{2}q}{\left(2\pi\right)^{2}}e^{i\vec{q}\cdot\vec{x}}D(q,0)$. Interestingly, the dynamics of the substrate now makes $D(q,0)\sim 1/q$ at low momenta (see Fig.~\ref{fig:height-height}a), since $\Delta_{0}(q,0)\propto 1/q$, while it would tend to a constant for a static substrate.
For large distances one therefore obtains $\left\langle h(x)h(0)\right\rangle \simeq k_{B}T/(2\pi K_{1}x)$, a result that is independent of $g$ and coincides with the result obtained for the surface out of plane displacement field of the bare substrate. For the quadratic dispersion
relation of flexural phonons, it is known that $\left\langle h(0)h(0)\right\rangle$ diverges at
any finite temperature, indicating the absence of crystalline
order. Coupling to a substrate makes this result finite, as can be seen in Fig.~\ref{fig:height-height}b. At high temperature  one can obtain the following approximate expressions:
\begin{equation}
\left\langle h(0)h(0)\right\rangle \simeq\begin{cases}
\frac{k_{B}T}{8\sqrt{\kappa g}} & ,\text{ for small }g\\
\frac{k_{B}T}{3\sqrt{3}\left(\kappa K_{1}^{2}\right)^{1/3}} & ,\text{ for large }g
\end{cases}.
\end{equation}
It is interesting to notice that for small $g$, the previous result coincides with the one that is obtained if one ignores the dynamics of a substrate \cite{Aranda_1998}.

\subsection{Thermal expansion}
Also of experimental interest is the areal  thermal
expansion coefficient, $\alpha_{A}$, of graphene on a substrate. It can be written in terms of the free energy $F=-k_{B}T\log Z$, with $Z$ the partition function, as
\begin{equation}
\alpha_{A}=-\frac{1}{B}\frac{\partial^{2}F}{\partial\Delta A\partial T},
\end{equation}
where $B=-A\left(\frac{\partial P}{\partial\Delta A}\right)_{T}=\lambda+\mu$ is the bulk modulus. Under an isotropic expansion of the membrane $\partial_{\alpha}u_{\beta}\rightarrow\bar{u}\delta_{\alpha\beta}+\partial_{\alpha}u_{\beta}$,
with a relative change of area given by $\Delta A/A=2\bar{u}$, the term $S_{\text{in}}$
will generate a new quadratic term in the action $S\supset-\int dtd^{2}x\bar{u}\left(\lambda+\mu\right)\left(\partial h\right)^{2}$. In a quasi-harmonic treatment we will keep this new term while still ignoring anharmonic terms in the action. Writing the partition function as a path integral in imaginary time,
$Z=\int D[h]e^{-S_{E}}$ with $S_{E}$ the Euclidean action \cite{Atland}, we can obtain $\alpha_{A}$ from
\begin{equation}
\alpha_{A}=\frac{1}{2}\frac{\partial}{\partial T}\left(\frac{k_{B}T}{A}\sum_{q,i\omega_{n}}q^{2}D(q,i\omega_{n})\right),
\end{equation}
where $D(q,i\omega_{n})$ is the Matsubara height-height Green's function.
Performing the Matsubara sum over bosonic frequencies $i\omega_{n}$ one obtains a generalization of the result found in Ref.~\onlinecite{Paco_2012}
\begin{equation}
\alpha_{A}=\frac{\hbar^{2}}{k_{B}T^{2}}\int\frac{d^{2}qd\omega}{\left(2\pi\right)^{3}}q^{2}\omega\frac{\text{Im}D(q,\omega)}{4\sinh^{2}\left(\frac{\hbar\omega}{2k_{B}T}\right)}.
\end{equation}
The obtained value is negative, since $\text{sgn}\left[\text{Im}D(q,\omega)\right]=-\text{sgn}(\omega)$,
and tends to a constant at high temperature, approximately given by
\begin{equation}
\alpha_{A}\simeq\begin{cases}
-\frac{k_{B}}{16\pi\kappa}\log\left(1+\frac{\kappa q_{D}^{4}}{g}\right) & ,\text{ for small }g\\
-\frac{k_{B}}{12\pi\kappa}\log\left(1+\frac{\kappa q_{D}^{3}}{K_{1}}\right) & ,\text{ for large }g
\end{cases},
\end{equation}
where $q_{D}$ is the Debye momentum. Close to room
temperature one obtains a value in the order of $-6$ to $-7\times10^{-6}\text{ K}^{-1}$,
a value that is smaller in absolute value than the one obtained for
a suspended membrane \cite{Paco_2012} (see Fig.~\ref{fig:exp_resist}a).

\subsection{Contribution to electrical resistivity}
\begin{figure}
\begin{centering}
\includegraphics[height=2cm]{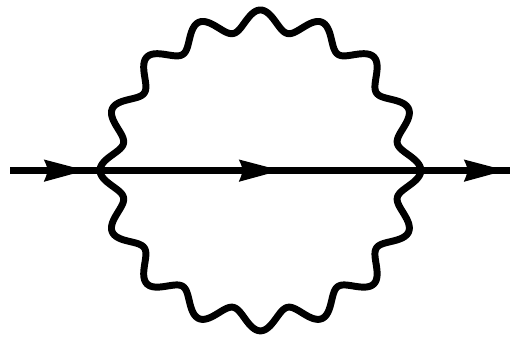}
\par\end{centering}

\caption{\label{fig:FeyDiagram}Second order sunset diagram contributing to
the electron self energy. The solid lines represent electron propagators,
while wiggly lines represent height-height membrane propagators.}

\end{figure}

\begin{figure}
\begin{centering}
\includegraphics[width=6.5cm]{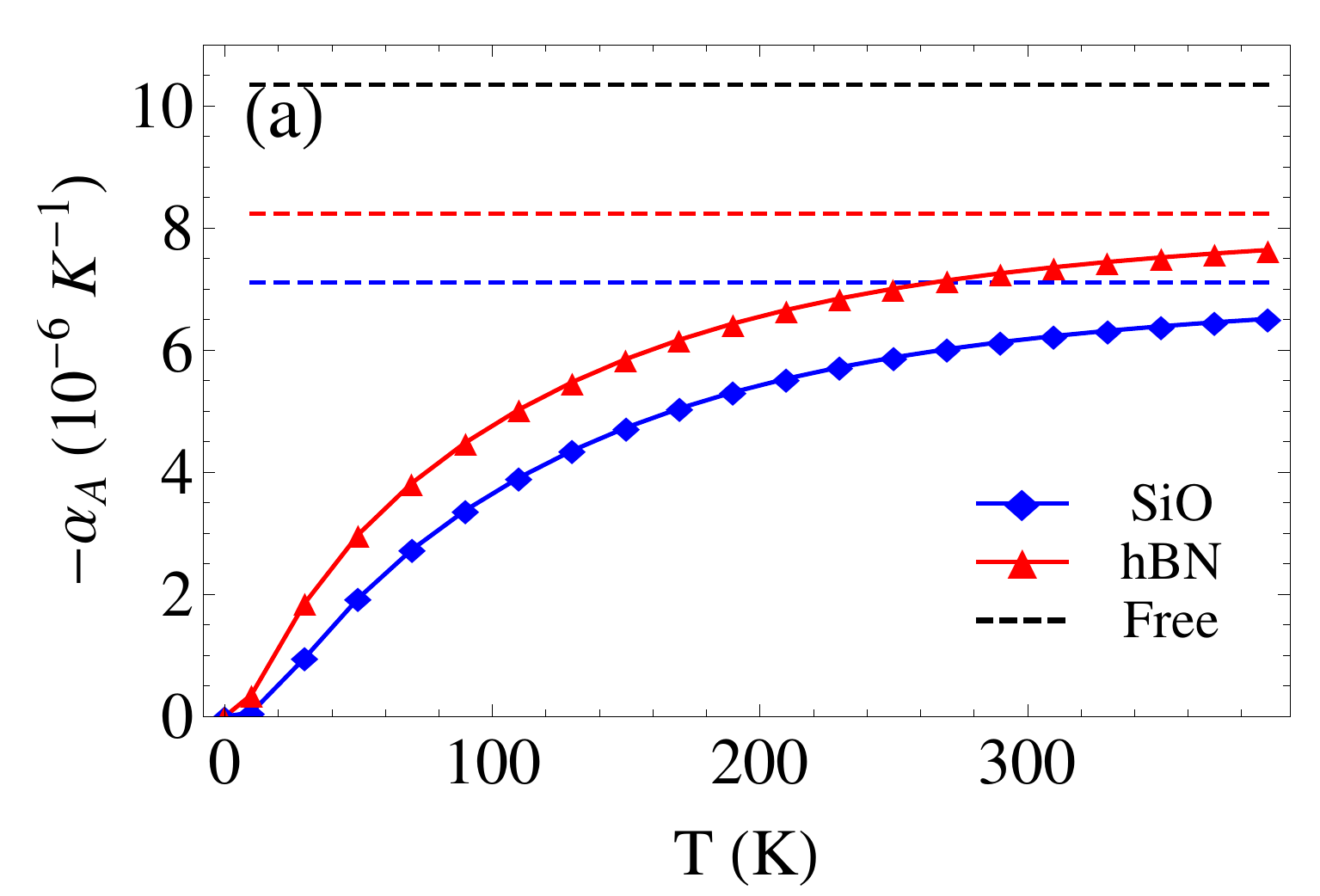}
\includegraphics[width=6.5cm]{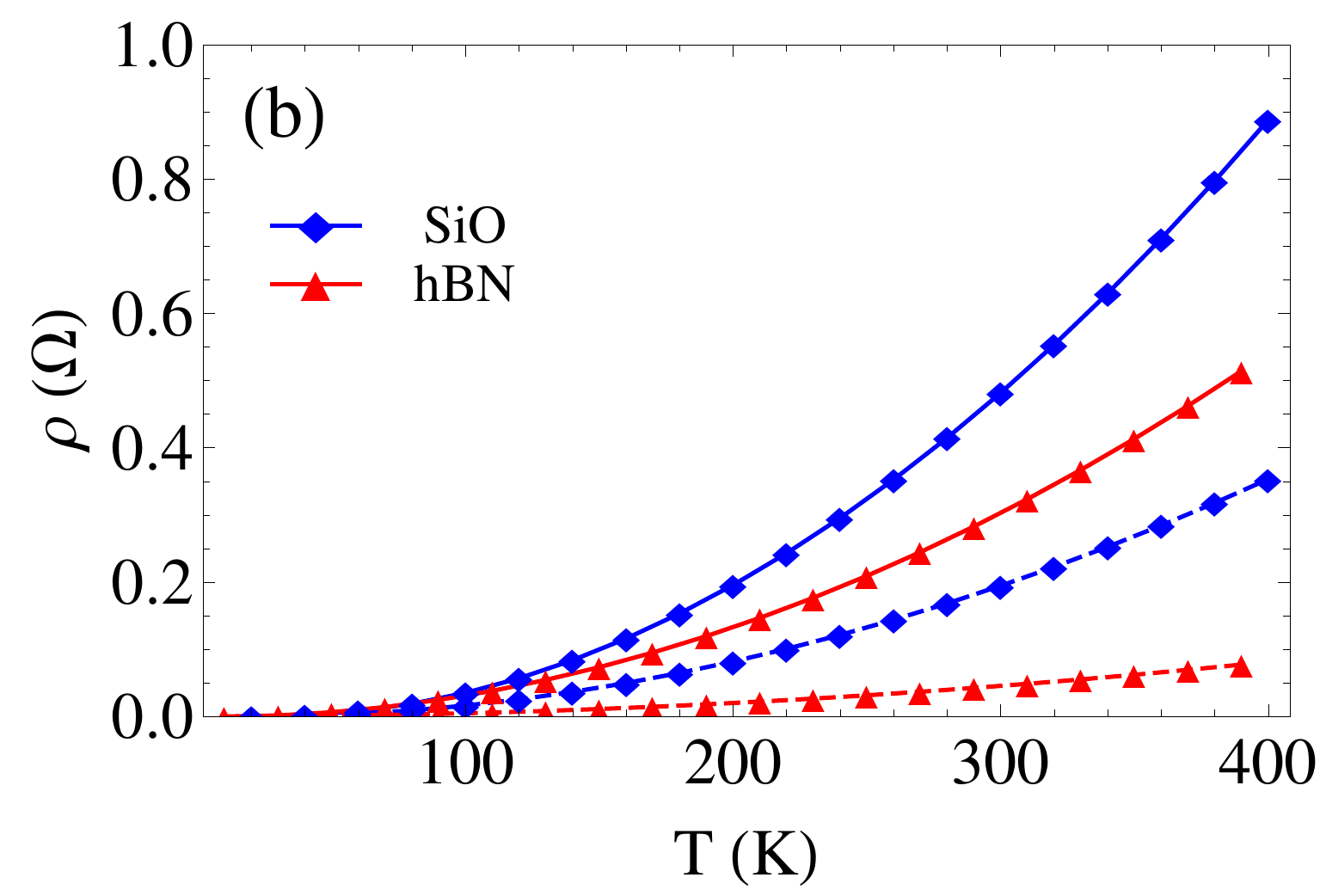}
\par\end{centering}

\caption{\label{fig:exp_resist}(a) Thermal expansion coefficient of graphene on SiO$_{2}$ and hBN substrates (solid lines) and high temperature
limits (dashed lines). Also shown is the estimated high temperature thermal expansion coefficient for free standing graphene (\onlinecite{Paco_2012}). (b) Electrical resistivity due
to flexural phonons in graphene on SiO$_{2}$ and hBN for a electronic density $n=10^{12}\text{ cm}^{-2}$. Dashed lines show the individual contribution of scattering by two fR modes. $g=1.82\times10^{20}\text{J}/\text{m}^{4}$ was used for both substrates.
}

\end{figure}

Knowing $D(q,\omega)$ one can also study the contribution
to the electrical resistivity of the flexural phonons on doped supported graphene. We compute the resistivity from the known formula $\rho^{-1}=\frac{e^{2}}{2}\mathcal{N}(\epsilon_{F})v_{F}^{2}\tau_{F}^{tr},$
where $\mathcal{N}(\epsilon_{F})=2k_{F}/\left(\pi\hbar v_{F}\right)$
is graphene density of states at the Fermi level, with $v_{F}$ and
$k_{F}$ the Fermi velocity and momentum respectively, and $\tau_{F}^{tr}$
the transport scattering time. In order to compute $\tau_{F}^{tr}$, one must describe the electron-phonon interaction in graphene. Assuming that graphene is electronically weakly coupled to the substrate,
the electron-phonon interaction in graphene should have the same form as
the one in free standing graphene \cite{Ando_2002,Manes_2007,RevModPhys_2009,Vozmediano_2010}:
\begin{align}
H_{\text{e-ph}}= & D_{0}\int d^{2}x\Psi^{\dagger}(x)\Psi(x)\varepsilon_{\alpha \alpha}(x)\nonumber\\
                 & -v_{F}\beta\int d^{2}x\Psi^{\dagger}(x)\vec{\sigma}\Psi(x)\cdot\vec{A}(x),
\end{align}
where $\Psi$ is the electron annihilation operator in the sublattice basis,  $D_{0}\approx 25\text{ eV}$ is the bare deformation potential, $\beta\approx 2.5\text{ eV}$ describes the change in electron hopping with bond stretching \cite{Ando_2002,Manes_2007,RevModPhys_2009,Vozmediano_2010}, $\vec{\sigma}=(\sigma_{x},\sigma_{y})$ is the 2D Pauli vector and $\vec{A}(x)$ is the vector potential induced by the distortion,
\begin{equation}
\vec{A}(x)=\frac{\hbar}{2a}\left(\varepsilon_{xx}-\varepsilon_{yy},\,2\varepsilon_{xy}\right).
\end{equation}
Notice that since the deformation potential is a coupling to the electronic density, it will be subject to screening \cite{EdCastro_2010,HOchoa_2011}. Focusing on the electron-flexural phonon interaction, after doing
a Fourier transform, writing the electron operator in the chiral basis
and focusing only in scattering in the conduction band (+), we can write
\begin{equation}
H_{\text{e-f}}=\frac{1}{2A^{2}}\sum_{k,q,p}w_{k,q,p}^{+,+}\psi_{+,k+q+p}^{\dagger}\psi_{+,k}h_{q}h_{p},
\end{equation}
with
\begin{align}
w_{k,q,p}^{+,+}= & -D_{0}qp\cos(\theta_{q,p})\frac{1}{2}\left(1+e^{i\theta_{k,k+q+p}}\right)\nonumber\\
                 & +\frac{\hbar v_{F}\beta}{2a_{0}}\frac{1}{2}qp\left(e^{i(\theta_{q}+\theta_{p}-\theta_{k^{\prime}})}+e^{-i(\theta_{q}+\theta_{p}-\theta_{k})}\right),
\end{align}
where $\theta_{k,k^{\prime}}=\theta_{k}-\theta_{k^{\prime}}$. To compute
the transport scattering time, $\tau_{k}^{tr}(\epsilon)$, we first
compute the electron lifetime, $\tau_{k}(\epsilon)$, in second order
in the electron-flexural phonon interaction. This can be obtained
from the imaginary part of the (retarded) self energy, computed from
the diagram in Fig.~\ref{fig:FeyDiagram}. Considering only scattering
in the conduction band one obtains
\begin{widetext}
\begin{equation}
\frac{1}{\tau_{k}(\epsilon)}=\frac{\pi}{\hbar A^{2}}\sum_{q,p}\int\frac{d\omega d\nu}{\pi^{2}}\frac{\hbar^{2}\left|w_{k,q,p}^{+,+}\right|^{2}\cosh\left(\frac{\epsilon-\epsilon_{F}}{2k_{B}T}\right)}{4\cosh\left(\frac{\hbar\omega+\hbar\nu+\epsilon-\epsilon_{F}}{2k_{B}T}\right)\sinh\left(\frac{\hbar\omega}{2k_{B}T}\right)\sinh\left(\frac{\hbar\nu}{2k_{B}T}\right)}\delta\left(\epsilon+\hbar\omega+\hbar\nu-\epsilon_{k+q+p}\right)\text{Im}D(q,\omega)\text{Im}D(p,\nu).
\end{equation}
To compute the transport scattering time, the sum in momentum must
be weighted by the factor $\left(1-\cos\theta_{k,k+q+p}\right)$ that
appears due to vertex corrections when computing the conductivity \cite{Mahan}.
In the quasi-elastic approximation for acoustic phonon
scattering one ignores the phonon energy in the energy conserving
Delta function and sets $\epsilon=\epsilon_{F}$, the electron Fermi
energy. We finally obtain a generalization of the result from Refs.~\onlinecite{EdCastro_2010,HOchoa_2011}
\begin{equation}
\frac{1}{\tau_{F}^{tr}}=\frac{\pi}{\hbar}\int\frac{d^{2}qd^{2}p}{\left(2\pi\right)^{4}}\int\frac{d\omega d\nu}{\pi^{2}}\hbar^{2}\left|w_{k_{F},q,p}\right|^{2}\left(1-\cos\theta_{k_{F},k_{F}+q+p}\right)d(\omega,\nu)\delta(\epsilon_{k+q+p}-\epsilon_{F})\text{Im}D(p,\nu)\text{Im}D(q,\omega),
\end{equation}
\end{widetext}
where $\left|w_{k,q,p}\right|^{2}=q^{2}p^{2}\mathcal{D}(\left|\vec{q}+\vec{p}\right|)^{2}$
is the squared electron-flexural phonon coupling, with $\mathcal{D}(Q)^{2}=D_{0}^{2}\left[1-Q^{2}/(4k_{F}^{2})\right]^{2}/\epsilon(Q)^{2}
+\hbar^{2}v_{F}^{2}\beta^{2}/(8a^{2}))$
the generalized deformation potential \cite{EdCastro_2010} and $\epsilon(Q)$ is the static dielectric function of graphene \cite{Stauber_2006}. We
have defined
$
d(\omega,\nu)=\frac{1}{4}\text{sech}\left(\frac{\hbar(\omega+\nu)}{2k_{B}T}\right)\text{csch}\left(\frac{\hbar\omega}{2k_{B}T}\right)\text{csch}\left(\frac{\hbar\nu}{2k_{B}T}\right)
$.
The electron scattering due to flexural phonons is a two phonon process.
Therefore there are three contributions: (i) scattering by two fG
modes, (ii) scattering by two fR modes, and (iii) a mixed process
with scattering by one fR and one fG mode. We see in Fig.~\ref{fig:exp_resist}b that the contribution from fR modes is smaller
in hBN than in SiO$_{2}$. This is explained by the smaller hybridization
with the Rayleigh mode in hBN due to a smaller Rayleigh velocity when
compared to SiO$_{2}$. We see that at room temperature
the obtained resistivities are of order $\sim1\Omega$, a value that is much smaller
than the expected contribution from the flexural mode in suspended samples ($\sim 200\Omega$) \cite{EdCastro_2010}
and the contribution from in plane phonons ($\sim 50\Omega$)~\cite{DasSarma_2008,EdCastro_2010,HOchoa_2011}.

\section{Conclusions}
We have modelled the dynamics
of the flexural mode of a membrane coupled to a dynamical substrate, with the aim
of understanding the role that the modified flexural mode might have in the
physics of graphene on a substrate. Since a half-space elastic medium supports both
3D bulk modes and a 2D surface Rayleigh mode, coupling of the membrane
modes to the substrate leads to a splitting of the spectral weight
of the flexural mode in two parts: one that will hybridize with the
substrate Rayleigh mode, acquiring an almost linear dispersion relation
for $q\lesssim c_{\text{R}}/\alpha$, and a second branch which resembles the original flexural mode, which becomes gapped and is broadened by the continuum of substrate bulk modes. This picture seems to be confirmed by experimental data \cite{Aizawa_1990PRB}.
As expected, coupling to the substrate leads to a stabilization of
the membrane and all correlation functions become finite, while for
a free membrane they are known to have infrared divergences in the harmonic
theory. It is worthwhile noticing that at high temperature the low
momentum behaviour of the height-height correlation function changes
from the $1/q^{4}$ of a free membrane to $1/q$ of the substrate. This implies that for large distances the height-height function will go as $1/x$. We also explored the behaviour of the areal
thermal expansion coefficient of graphene on substrate. At room temperature
we obtained a value of the order of $-6$ to $-7\times10^{-6}\text{ K}^{-1}$.
Finally we studied the contribution of flexural modes on the electrical
resistivity of doped graphene supported by a substrate. We found that coupling of the membrane to the
substrate strongly suppresses the contribution of the flexural
phonons, even if one takes into account the contribution coming from
the hybridized flexural-Rayleigh mode. Note that the model describes a flat graphene layer on a flat surface. This is a good approximation for substrates such as hBN.  In the case of corrugated substrates, such as SiO$_2$, our analysis is expected to describe the regions where the graphene layer and the substrate are flat and the two systems are in close contact.

\begin{acknowledgments}
The authors would like to thank J. Schiefele, H. Ochoa, E. Cappelluti, R. Roldan, N.M.R. Peres and M.I. Katsnelson for useful discussions. B.A. acknowledges financial support from Funda\c{c}\~{a}o para a Ci\^{e}ncia e a Tecnologia, Portugal, through Grant No. SFRH/BD/78987/2011. F.G. acknowledges financial support from MINECO, Spain, through grant FIS2011-23713, and the European Research Council Advanced
Grants program, through grant 290846.
\end{acknowledgments}

\appendix*
\section{Elastic response of uniaxial substrate}
\label{appendix}
In a quadratic theory, the quantum mechanical retarded Green's function
coincides with the classical Green's function. Therefore, in order
to determine the Green's function of a semi-infinite elastic medium,
\begin{equation}
\Delta_{0}^{ij}(\boldsymbol{x},t;\boldsymbol{x}^{\prime},t^{\prime})=-\frac{i}{\hbar}\Theta(t-t^{\prime})\left\langle \left[u_{i}^{(s)}(\boldsymbol{x},t),\, u_{j}^{(s)}(\boldsymbol{x}^{\prime},t^{\prime})\right]\right\rangle
\end{equation}
(where $\boldsymbol{x}=(\vec{x},z)$), we study the classical response of the substrate to an external pressure, $\vec{\sigma}(\vec{x},t)$,
at the boundary $z^{\prime}=0$:
\begin{equation}
u_{i}^{(s)}(\vec{x},z,t)=-\int dt^{\prime}d^{2}x^{\prime}\Delta_{0}^{ij}(\vec{x}-\vec{x}^{\prime},z,t-t^{\prime})\sigma_{j}(\vec{x}^{\prime},t^{\prime}),
\end{equation}
where $\Delta_{0}^{ij}(\vec{x}-\vec{x}^{\prime},z,t-t^{\prime})=\Delta_{0}^{ij}(\vec{x},z,t;x^{\prime},0,t^{\prime})$ and we have used the fact that there is translational invariance in the $\vec x$-plane and in time.

The displacement field obeys the bulk
equations of motion 
\begin{equation}
\left(-\partial_{t}^{2}\delta_{ij}+c_{iklj}\partial_{k}\partial_{l}\right)u_{j}^{(s)}(\vec{x},z,t)=0
\end{equation}
and the boundary conditions
\begin{equation}
c_{3ijk}\partial_{j}u_{k}^{(s)}(\vec{x},0,t)=\sigma_{i}(\vec{x},t).
\end{equation} 
Obtaining the solution for $u_{i}^{(s)}$ in the presence of the external pressure, we can read of $\Delta_{0}^{ij}$.
We look for solutions of the form $\vec{u}^{(s)}(\vec{x},z,t)=\int\frac{dtd^{2}q}{(2\pi)^{3}}\vec{u}^{(s)}(\vec{q},\omega,z)e^{i\left(\vec{q}\cdot\vec{x}-\omega t\right)}$,
with $\vec{u}^{(s)}(\vec{q},\omega,z)=\sum_{\lambda=1}^{3}a_{\lambda}(\vec{q},\omega)\vec{\xi}_{\lambda}(\vec{q},\omega)e^{ip_{\lambda}(\vec{q},\omega)z}$,
where $p_{\lambda}(\vec{q},\omega)$ and $\vec{\xi}_{\lambda}(\vec{q},\omega)$
are determined by the bulk equations of motion and the coefficients
$a_{\lambda}(\vec{q},\omega)$ are fixed by the boundary conditions.
In order to obtain a retarded response, we must pick the solutions
for $p_{\lambda}(\vec{q},\omega)$ such that the real part has an
opposite sign from $\omega$; and to obtain a finite response we pick
the solutions that have a positive imaginary part. We are interested in the response of the substrate at the boundary $z=0$. This can be written in matrix form as $u_{i}{(s)}(\vec{q},\omega,0)=-\Delta_{0}^{ij}(\vec{q},\omega)\sigma_{j}(\vec{q},\omega)$, from which one can read the coefficient $\Delta_{0}^{zz}(\vec{q},\omega)$ which was simply written as $\Delta_{0}(q,\omega)$ in Sec~\ref{sec:Model}.

An uniaxial material has a plane of isotropy, having $5$ independent elastic constants.
In Voigt notation, the elastic constants tensor is given by:
\begin{equation}
c_{IJ}=\left[\begin{array}{cccccc}
c_{11} & c_{12} & c_{13}\\
c_{12} & c_{11} & c_{13}\\
c_{13} & c_{13} & c_{33}\\
 &  &  & c_{44}\\
 &  &  &  & c_{44}\\
 &  &  &  &  & \left(c_{11}-c_{12}\right)/2
\end{array}\right].
\end{equation}
We can set $\vec{q}=(0,q)$ without loss of generality. The final result from the calculation is given by
\begin{align}
\Delta_{0}^{xx}(q,\omega) & =\frac{i}{c_{44}p_{3}},\nonumber \\
\Delta_{0}^{xi}(q,\omega) & =\Delta_{0}^{ix}(q,\omega)=0,\, i=y,z,\nonumber \\
\Delta_{0}^{yy}(q,\omega) & =iM^{-1}c_{33}\left(f_{1}p_{2}-f_{2}p_{1}\right),\nonumber \\
\Delta_{0}^{yz}(q,\omega) & =iM^{-1}c_{44}\left[f_{1}f_{2}(p_{1}-p_{2})-q(f_{1}-f_{2})\right],\nonumber \\
\Delta_{0}^{zy}(q,\omega) & =iM^{-1}\left[c_{33}(p_{2}-p_{1})+c_{13}q(f_{2}-f_{1})\right],\nonumber \\
\Delta_{0}^{zz}(q,\omega) & =iM^{-1}c_{44}\left(f_{1}p_{1}-f_{2}p_{2}\right),
\end{align}
where we have defined
\begin{align}
p_{3} =   & -\text{sgn}(\omega)\sqrt{\frac{\omega^{2}\rho_{\text{3D}}}{c_{44}}-\frac{c_{11}-c_{12}}{2c_{44}}q^{2}+\text{sgn}(\omega)i0^{+}},\nonumber\\
p_{1/2} = & -\text{sgn}(\omega)\sqrt{\frac{1}{2}B\pm\frac{1}{2}\sqrt{B^{2}-4C}+\text{sgn}(\omega)i0^{+}},\nonumber \\
f_{1/2} = & \frac{(c_{13}+c_{44})qp_{1/2}}{\omega^{2}-c_{11}q^{2}-c_{44}p_{1/2}^{2}},\nonumber\\
M =       & c_{44}\left(p_{1}f_{1}+q\right)\left(c_{13}qf_{2}+c_{33}p_{2}\right)\nonumber\\
          & -c_{44}\left(p_{2}f_{2}+q\right)\left(c_{13}qf_{1}+c_{33}p_{1}\right),
\end{align}
with
\begin{align}
B & = \frac{c_{11}}{c_{44}}\left(\frac{\omega^{2}\rho_{\text{3D}}}{c_{11}}-q^{2}\right)+\frac{c_{44}}{c_{33}}\left(\frac{\omega^{2}\rho_{\text{3D}}}{c_{44}}-q^{2}\right)+\frac{\left(c_{13}+c_{44}\right)^{2}}{c_{33}c_{44}}q^{2}\nonumber \\
C & = \frac{c_{11}}{c_{33}}\left(\frac{\omega^{2}\rho_{\text{3D}}}{c_{11}}-q^{2}\right)\left(\frac{\omega^{2}\rho_{\text{3D}}}{c_{44}}-q^{2}\right).
\end{align}

The condition to have a surface Rayleigh mode is determined by $M=0$.

In the case of an isotropic substrate, $c_{11}=c_{33}=\lambda_{\text{3D}}+2\mu_{\text{3D}}=\rho_{\text{3D}}c_{\text{L}}^{2}$,
$c_{12}=c_{13}=\lambda_{\text{3D}}=\rho_{\text{3D}}(c_{\text{L}}^{2}-2c_{\text{T}}^{2})$,
$c_{44}=\mu_{\text{3D}}=\rho_{\text{3D}}c_{\text{T}}^{2}$, one recovers the result
from Ref.~\onlinecite{Persson_2001}. In particular $\Delta_{0}(q,\omega)=\Delta_{0}^{zz}(q,\omega)$ reads:
\begin{equation}
\Delta_{0}(q,\omega)=-\text{sgn}(\omega)\frac{i\omega^{2}}{c_{\text{T}}^{4}S(q,\omega)}\sqrt{\left(\frac{\omega}{c_{\text{L}}}\right)^{2}-q^{2}+\text{sgn}(\omega)i0^{+}},
\end{equation}
with
\begin{widetext}
\begin{equation}
S(q,\omega)=\left(\left(\frac{\omega}{c_{\text{T}}}\right)^{2}-2q^{2}\right)^{2}+4q^{2}\sqrt{\left(\frac{\omega}{c_{\text{T}}}\right)^{2}-q^{2}+\text{sgn}(\omega)i0^{+}}\sqrt{\left(\frac{\omega}{c_{\text{L}}}\right)^{2}-q^{2}+\text{sgn}(\omega)i0^{+}}.
\end{equation}
\end{widetext}


%

\end{document}